\documentclass[prl,preprint,showpacs,byrevtex,endfloats]{revtex4}

\usepackage{graphicx}
\usepackage{epsfig}
\preprint{Daukiya et al.}
\usepackage{amssymb}
\usepackage{amsmath}
\usepackage{amstext}
\usepackage[french,english]{babel}
\usepackage[applemac]{inputenc}
\begin{document}
\def\BibTeX{\rm B{\sc ib}\TeX}

\title{Covalent functionalization by cycloaddition reactions of pristine, defect-free graphene\\}
\author{L. Daukiya $^{1}$, C. Mattioli$^{2}$, D. Aubel$^{1}$, S. Hajjar-Garreau$^{1}$, F. Vonau$^{1}$, E. Denys$^{1}$, G. Reiter$^{3}$, J. Fransson$^{4}$, E. Perrin$^{5}$, M-L. Bocquet$^{5}$, C. Bena$^{6-7}$, A. Gourdon$^{2}$ and L.
Simon$^{1}$\footnote[1]{corresponding author \\Email address: laurent.simon@uha.fr}}
\affiliation{$^{1}$Institut des Sciences des Mat\'eriaux de Mulhouse, CNRS-UMR 7361, Universit\'e de
Haute Alsace, 3Bis, rue Alfred Werner, 68093 Mulhouse, France.}

\affiliation{$^{2}$ Nanosciences group, CEMES CNRS-UPR 8011, 29 Rue Jeanne Marvig, BP 94347, 31055 Toulouse, France}
\affiliation{$^{3}$Physikalisches Institut, Universit\"at Freiburg, Hermann-Herder-Strasse 3, 79104
Freiburg, Germany}
\affiliation{$^{4}$Department of Physics and Astronomy, Uppsala University, Box 516, SE-751 21 UPPSALA, Sweden}
\affiliation{$^{5}$ Dpt of Chemistry, UMR ENS-CNRS-UPMC 8640, Ecole Normale Superieure, F-75005 Paris, France}
\affiliation{$^{6}$Institut de Physique Th\'eorique, CEA/Saclay, Orme des Merisiers, 91190 Gif-sur-Yvette Cedex, France}
\affiliation{$^{7}$Laboratoire de Physique des Solides, CNRS, UMR-8502. Paris Sud, 91405 Orsay CEDEX, France}

\date{\today}

\begin{abstract}

Based on a low temperature scanning tunneling microscopy study, we present a direct visualization of a cycloaddition reaction performed for some specific fluorinated maleimide molecules deposited on graphene. These studies showed that the cycloaddition reactions can be carried out on the basal plane of graphene, even when there are no pre-existing defects. In the course of covalently grafting the molecules to graphene, the $sp^{2}$ conjugation of carbon atoms was broken  and  local $sp^{3}$ bonds were created. The grafted molecules perturbed the graphene lattice, generating a standing-wave pattern with an anisotropy which was attributed to a (1,2) cycloaddition, as revealed by T-matrix approximation calculations. DFT calculations showed that while both (1,4) and (1,2) cycloaddition were possible on free standing graphene, only the (1,2) cycloaddition could be obtained for graphene on SiC(0001). Globally averaging spectroscopic techniques, XPS and ARPES, were used to determine the modification in the elemental composition of the samples induced by the reaction, indicating an opening of an electronic gap in graphene.

\end{abstract}


\maketitle

\begin{figure}
\begin{center}
\includegraphics[width=6in]{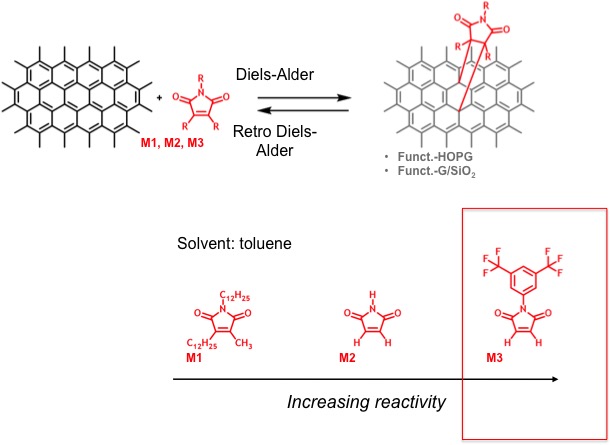}
\end{center}
\caption{Description of a Diels-Alder reaction as an example of cycloaddition on pristine graphene. Here, graphene is considered as a diene which reacts with a dienophile molecule. We used 3,5- bis(trifluoromethyl)phenyl substituted maleimide derivatives (named  fluorinated maleimide (FMAL)). Several molecules of the maleimide-type M1 to M3 have been tested. Only M3-type molecules reacted with  epitaxial graphene. The fluorinated atoms are introduced to polarise the double bond of the maleimide group, and so strongly increased the dienophilic character of the molecule \cite{Mattioli}. }\label{Fig1}
\end{figure}

Because of its fascinating properties, graphene is presently considered for a large number of potential  applications. However, functionalizing this gapless highly non-reactive semiconductor with the appropriate molecules remains an important challenge.  Combined with supramolecular chemistry, chemical functionalization, i.e., creating covalent bonds by converting sp$^2$ into sp$^3$ orbitals, represents a  promising path, because it allows to selectively  modify the surface of graphene with a high spatial control.  Such a surface modification (hybridization) can be realized by Diels-Alder or other cycloaddition reactions.

Graphene undergoes cycloaddition or Diels Alder (D-A) reactions mainly because of the degeneracy of the electronic states at the Dirac point. The states close to the Fermi level may give rise to either an anti-symmetric or symmetric graphene orbital. These orbitals allow graphene to function as both donor and acceptor within the Frontier Molecular Orbital (FMO) theory \cite{HaddonJACS2011, HaddonACR2012, Fukui1970, TownshendJACS1976}. For exfoliated graphene, a D-A reaction and its reversibility  have been demonstrated for the first time by the team of Robert C. Haddon \cite{HaddonJACS2011}. The progress of the D-A reaction was followed by Raman spectroscopy and the ratio between the G and D bands which ascertained that graphene was modified. However, such global measurements  did not allow to identify the locations where graphene was actually functionalized. This seminal work by Haddon et al. raises the question if pre-existing defect (step-edges, holes,...) are required to allow for the D-A reaction. Subsequently,  two theoretical studies \cite{HoukJACS2013, DenisChemEur2013} concluded that only the edges of  graphene layers, or holes therein, may be functionalized by a cycloaddition reaction. However, no such reaction was predicted to be possible within the bulk of a clean graphene layer. Indeed,  calculation results  for a D-A reaction in the bulk predict a highly endothermic value of up to 2.6 eV. Recently, by combining scanning tunneling microscopy (STM) and density functional theory (DFT) calculations, a cycloaddition reaction has been evidenced for a  graphene layer on iridium, involving a non-expected 1,3 cycloaddition of graphene in the low electron density regions of the Moir\'e pattern \cite{Altenburg2015}. There, the allyl reactivity of epitaxial graphene was due to the presence of the substrate Ir.  The endothermic character of the cycloaddition reaction was preserved, but was characterized by a more modest value of + 0.4 eV, accessible by choosing an appropriately high temperature. This observation demonstrates that  a D-A reaction is possible when using a graphene layer 'activated'  by an underlying  metal substrate.

Here, we combine STM, angle resolve photoemission spectroscopy (ARPES) and X-ray photoelectron spectroscopy (XPS) to study the
Diels-Alder reaction on a monolayer (ML) and a bilayer (BL) epitaxial graphene on SiC(0001), using recently synthesized fluorinated maleimide molecules (see Figure\ref{Fig1}). The maleimide group of these molecules represents the reactive part, and the fluorine atoms were introduced to polarize the double bonds of the maleimide group and thereby  strongly increased the dienophilic character of the molecule and thus its reactivity \cite{Mattioli}. To deposit these molecules onto graphene, the substrates were simply immersed for several tens of hours in a toluene solution containing the molecules. After retraction from the solution, the samples were  intensively  rinsed. Our results provide strong evidence of the formation of chemical bonds between these molecules and the graphene layer through the cycloaddition reaction . The formation of the chemical bonds was ascertained by the increase of the $sp^{3}$ component of the C1s peak, and a decrease of the Fermi velocity. In addition, a tendency for the opening of a gap was evidenced by ARPES.

Moreover, the STM images provide also a strong evidence for the formation of covalent bonds by a  cyclo-addition reaction in the middle of clean graphene terraces. Indeed, chemically grafting these molecules to the graphene layer generates standing waves pattern which indicates that graphene is locally highly perturbed. The grafted molecules are associated to bright features on the STM images that are dispersed throughout the bulk of graphene terraces where no defect is expected in pristine graphene, at least not with such a high density. The smallest of these features are surrounded by standing waves which take the form of parallel straight lines, similar in anisotropy to those observed close to armchair graphene step edges. We present an experimental comparison of these standing-wave patterns with patterns generated by other possible defects, as well as with patterns generated by armchair and zigzag step edges. We also discuss how identifying the type of dominant quasiparticle scattering may reveal the nature of the cycloaddition reaction. Thus we use T-matrix approximation calculations to compare quasiparticle interference patterns and the local density of states when assuming various possible configurations of atoms affected by the cycloaddition reaction, in particular a one atom (1,1) configuration, and some two-atom configurations [(1,2), (1,3), or (1,4)] . We find that the shape of the standing-wave patterns depends not only on the defect geometry, but also on the type of sublattice combinations  (AA or AB) of the atoms affected by the reaction. The highly anisotropic perturbation pattern of the density of states was reproduced for the (1,2) and (1,4) configurations of the cycloaddition reaction. Moreover, we show via large-scale DFT calculations on realistic graphene layers on a SiC substrate that  only  the (1,2) cycloaddition is stable, but  not the (1,4) or (1,3) ones. On bilayer graphene, the STM revealed similar patterns; moreover using XPS we measured an increase of the $sp^{3}$ component in the peak corresponding to the C1s core level, and the band structure modifications were significantly more pronounced.\\
In this study, we used   maleimide derivative molecules. As depicted in figure \ref{Fig1}, when interacting with these molecules in a D-A reaction, graphene can be considered as a diene   reacting with a dienophile molecule. However, at this stage it was not possible to predict which type of cycloaddition reaction will occur. This is particularly difficult for  graphene on SiC(0001) which is n-doped (thus the chemical reaction does not occur at the Dirac point). Specifically, we used 3,5- bis(trifluoromethyl)phenyl substituted maleimide derivatives (for which we use the abbreviated name:  fluorinated maleimide (FMAL)). Several  maleimide-type molecules (M1 to M3) have been tested. The reactivity of the molecules has been studied for  graphene on SiO$_2$ and graphene on SiC(0001) using Raman spectroscopy\cite{Dujardin2015}. It was found that only M3-type molecules reacted with  epitaxial graphene.
We demonstrate here that we can chemically functionalize graphene, a fascinating possibility which opens up the road for future research in the functionalization of graphene, as well for more systematic studies in particular using higher spatial resolution investigations. 

\begin{figure}
\begin{center}
\includegraphics[width=6in]{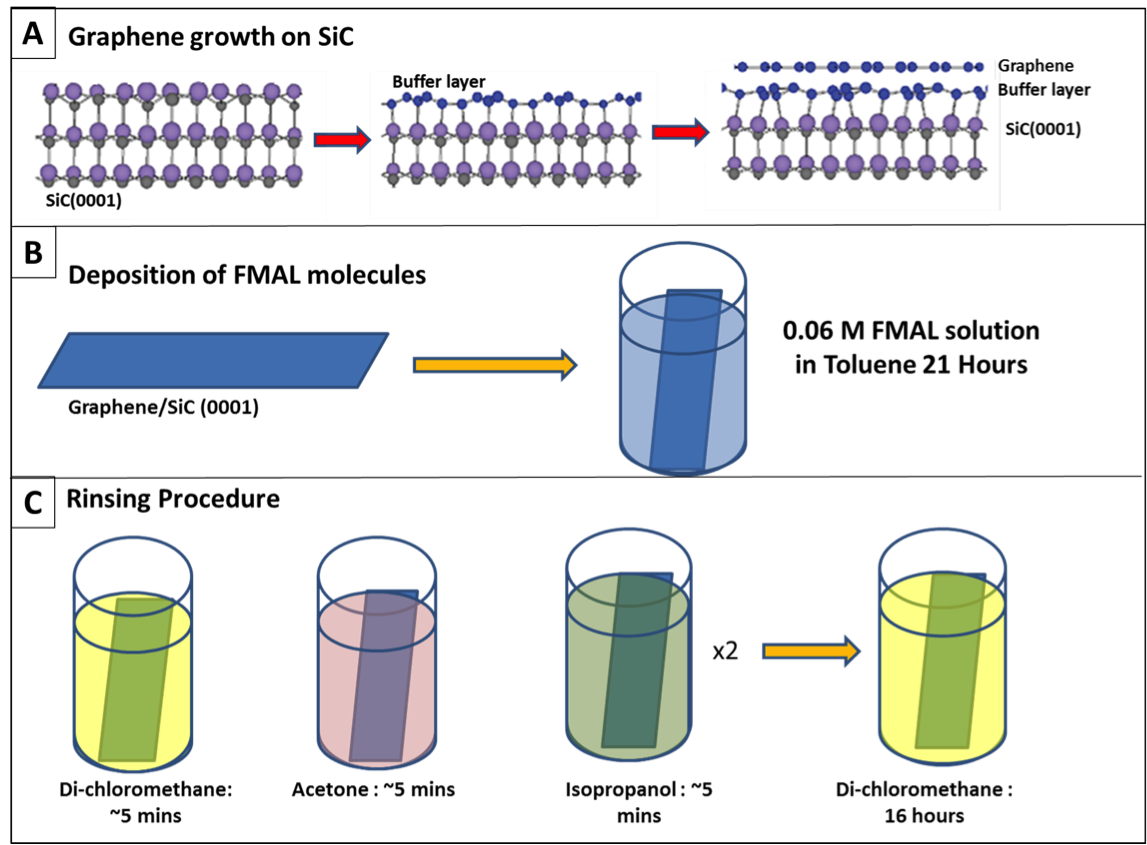}
\end{center}
\caption{. (A). Description of the growth of Graphene on SiC. Graphene was generated by exodiffusion and evaporation of silicon atoms from the SIC substrate by annealing at appropriate temperatures. This process lead to formation of covalently bonded layer of carbon atoms called buffer layer, and the subsequent formation of one or more graphene layers. (B) Monolayer or Bilayer graphene samples were dipped in the solution of FMAL molecules in toluene for different duration. (C) shows the rinsing procedure for removal of physisorbed molecules}\label{Fig2}
\end{figure}

\section{EXPERIMENTAL METHOD}
 
The scheme in figure \ref{Fig2} illustrates the growth of graphene on SiC (0001) by thermal evaporation of Si atoms. A SiC(0001) substrate was  annealed at temperatures above $1250^{\circ}$C \cite{simonPRB99,berger2002}. This led  to the exodiffusion of silicon atoms and the formation of a buffer layer which  is a graphene monolayer partially covalently bonded to the silicon atoms of the SiC(0001) substrate exhibiting a SiC-6x6 superstructure. This buffer layer is semiconducting. The graphene monolayer (ML graphene) or graphene bilayer (BL graphene) showed the characteristic properties of graphene and interacted via van der Waals forces with the buffer layer. We controlled the number of graphene layers via the annealing time and annealing  temperature. To obtain homogeneous surfaces, we assured a low pressure (less than $1.10 ^{-10} $mbar) during the annealing. The resulting graphene layers were characterized by  XPS and ARPES measurements. We used an hemispherical electron analyzer (Scienta R3000), and a monochromatic  X-ray source ($AlK_{\alpha}$). For ARPES we used the UV source HeII (40.6eV). Our STM experiments were performed at 77 K at a base pressure in the $10^{-11}$ mbar range in ultra high vacuum (UHV) with a LT-STM from Omicron.
Following the procedure represented in fig .\ref{Fig2}, the deposition of molecules was carried out on monolayer as well as
bilayer graphene.

\section{THEORETICAL METHOD}
The DFT calculations were carried out using the Vienna {\it ab initio} simulation  package (VASP)\cite{kresse1996efficient, kresse1996efficiency}, with the generalized-gradient approximation of PBE type \cite{ perdew1996generalized} as the XC functional. The electron-ion interaction was described by the projector augmented wave method \cite{blochl1994projector,kresse1999ultrasoft}. Plane waves were used  as the basis set, and the energy and  augmentation charge cutoffs were set to 300 eV and 645 eV. In order to get a detailed and realistic understanding of the  graphene/SiC interfacial structure on the Si-face of our SiC substrates, and to derive its influence to the electronic structure of graphene, we adopted an adequately large area of commensurability, i.e., a $13\times13$ graphene domain on a $6\sqrt{3}\times 6\sqrt{3}$  SiC substrate (referred as to 6R3) \cite{charrier_solid-state_2002,riedl_structural_2007,brar_scanning_2007}. The  SiC substrate was modeled by two SiC bilayers, the top face being Si and the bottom being C, saturated by hydrogen atoms, in total including 432 atoms. All structures were relaxed until the total forces were smaller than 0.05 eV/\AA. On this substrate, different types of cycloaddition reactions with the fluorinated maleimide molecule have been calculated. The stabilized cycloadducts have been compared with respect to structure and  energy with the ones formed on free-standing graphene.  \\

\section{RESULTS}
\begin{figure}
\begin{center}
\includegraphics[width=6in]{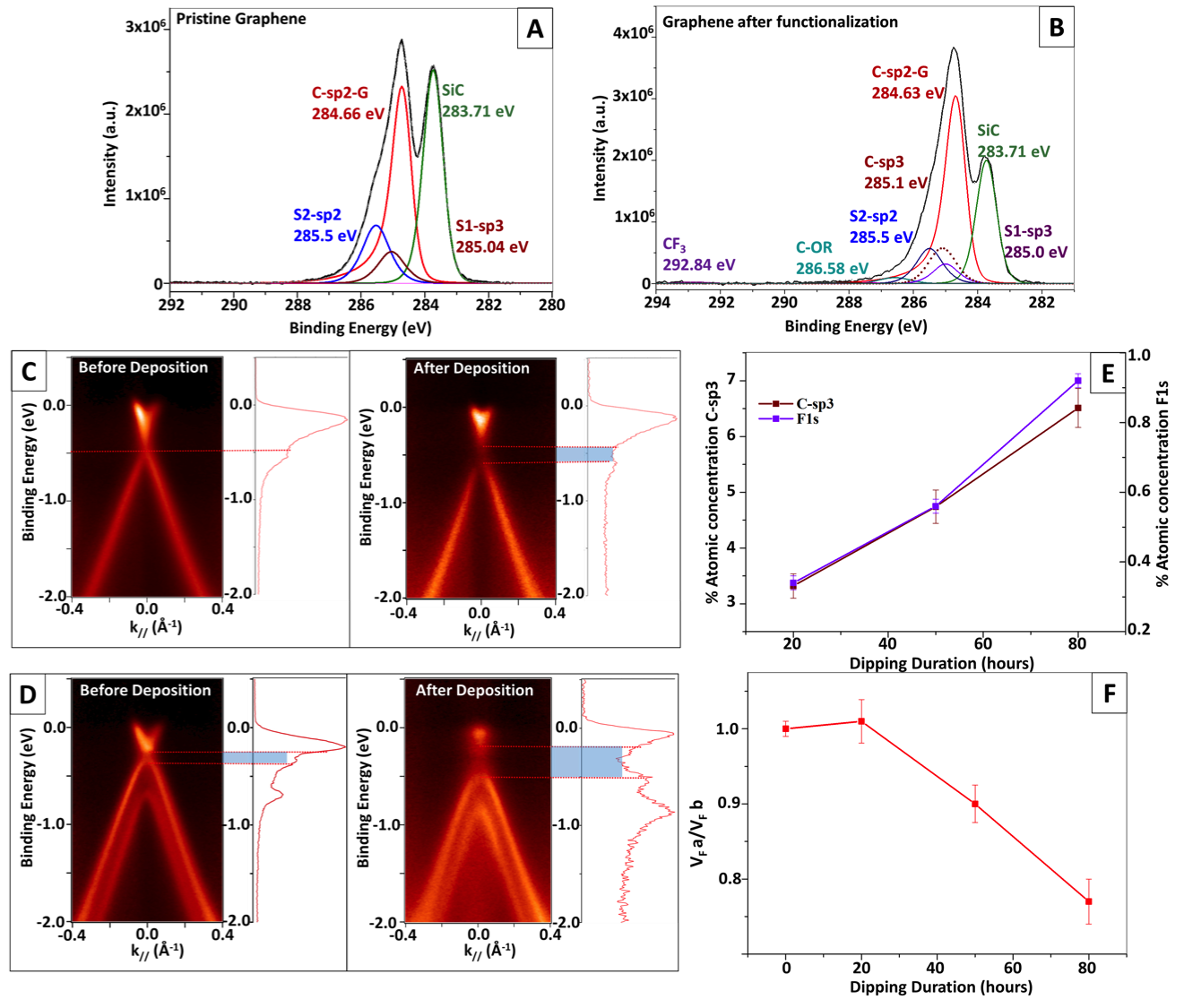}
\end{center}
\caption{Global measurements (XPS and ARPES) of graphene before and after the reaction with FMAL molecules (dipping duration 80 hours). A) and B) shows the deconvoluted C1s spectra before and after dipping respectively. The successful reaction of FMAL molecules is represented by the change in sp2 component which is associated with graphene and sp3 (dotted lines). In the supplementary information, we present the F1s, N1s and O1s components which assure that the molecule is intact. E) shows the evolution of sp3 component of C1s spectra and F1s spectra as a function of dipping time. C) and D) shows the band structure measured by ARPES for monolayer and bilayer graphene before and after the DA reactions, respectively. The constant momentum dispersion curve shows a dip in the intensity as an a tendency for opening of a gap. The change in the Fermi velocity for ML graphene as a function of the immersion time is given in F).}\label{Fig3}.
\end{figure}

Figure \ref{Fig3} A) shows a typical XPS spectrum of a pristine ML graphene. We can identify four components. The C1s component of SiC  corresponds to the carbon atoms of the SiC substrate. The $C-sp^{2}$ component is attributed to the graphene top layer and the $S1-sp^{3}$ and $S2-sp^{2}$ components correspond to the buffer layer carbon atoms covalently bonded to silicon atoms and carbon atoms in the graphene-like structure of the buffer layer, respectively \cite{Emtsev2008, RiedlJPhysD2010}. Figure \ref{Fig3} B) shows the corresponding peaks for the same sample after having been immersed in a  FMAL toluene solution for 80 hours,  followed by the  rinsing  procedure described in figure \ref{Fig2}.  The presence of FMAL molecules on the surface of graphene  resulted in a marked C-$sp^2$ (G) component and emergence of a new C-sp$^{3}$ component shown as dotted line in figure\ref{Fig3}B). The attachment of  molecules via their aromatic ring contributed to the increase of the component G. C-sp$^{3}$ is associated to the graffting of the molecules, this component emerges in the deconvolution by keeping constant the ratio of S1-$sp^{3}$ and S2-$sp^{2}$ with the SiC component before and after the immersion which leads to consider that the buffer layer is unchanged in the process.

The formation of $sp^{3}$ bonds, expressed by the increase of the component S1, was associated with the  grafting of the molecules. The other contributions of carbon configurations were less important in intensity but  still could be identified. Typically, as we will demonstrate for a cycloaddition for the  (1,2) configuration,  grafting of a single  molecule changed the $sp^{2}$ orbitals of 5 bonds into $sp^{3}$ orbitals.  Even for longer immersion times, due to the low sensitivity factor of the organic component, the contributions of grafted molecules in the C1s spectra were much less intense and contributed mainly to the tail of the C1s peak, i.e. for binding energies above 288 eV (see figure D) in supporting information). As provided in the supplementary information, the O1s, N1s and F1s spectra and their respective positions showed unambiguously that the FMAL molecules were deposited in the intact form  on the graphene surface.  In  Figure \ref{Fig3}E), we present a  plot of the intensity of the S1 component,  together with the intensity of F1s,  for immersion times ranging from 20 to 80 hours. Both intensities exhibited the same evolution, assuring  that the increase in $sp^{3}$ bonds could be  attributed to an increase in number of grafted molecules.

Figures \ref{Fig3} C) and D) show the dispersion bands measured by ARPES around the K-point. The band structure was measured before and after deposition of molecules on monolayer graphene (Figure \ref{Fig3} C) and bilayer graphene (Figure \ref{Fig3} D), respectively.
 We observe that the band structure was preserved and could  clearly be measured by ARPES, even after immersion and without any further treatment of the sample in UHV.  A tendency for the opening of  a small gap was marked by the decrease of the intensity around the Dirac point which is much more pronounced than for pristine graphene. This is clearly visible on the profiles taken as a function of energy at constant momentum at the K point (see figure \ref{Fig3}). In addition, the Fermi velocity, measured via the slope of the linear dispersion, decreased with  immersion time. This indicates that the number of covalently bonded molecules increased. A similar behavior has been observed, for example, for  induced defects caused by ion bombardment \cite{TapasztoPRB08}. We notice  that the opening of the gap  was  more pronounced for the case of bilayer graphene shown in Figure \ref{Fig3} D). The dispersion of the lower energy band, which corresponds to the upper graphene layer, and that initially was linearly crossing the Dirac point became rather parabolic. This change revealed an increase of the effective mass, as expected, and is consistent with the opening  of a gap.

\begin{figure}
\begin{center}
\includegraphics[width=6in]{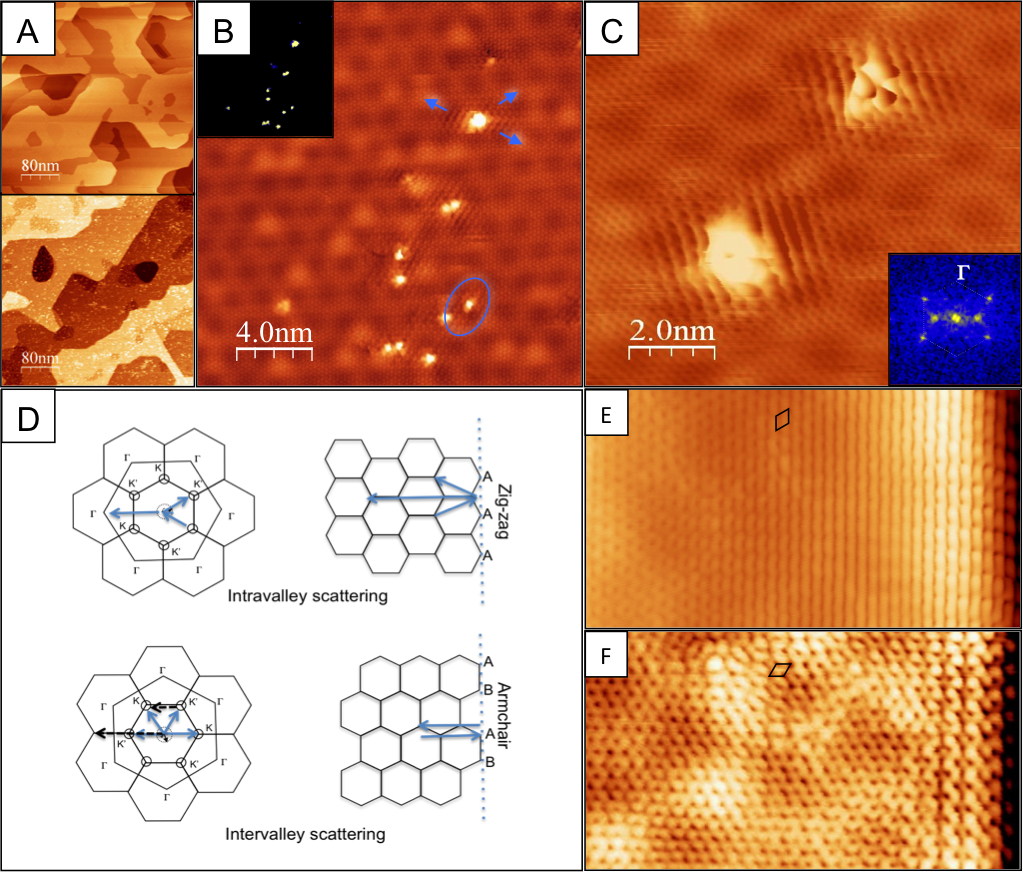}
\end{center}
\caption{ A) Large area (400x400 $nm^{2}$) STM scan (-1.5 V, 500 pA) on a pristine, defect-free graphene layer, as loaded in UHV, together with the corresponding image (-1.5V, 518 pA) for this sample   after immersion for  80 hours in a toluene solution of FMAL molecules. (B) 20x20 $nm^{2}$ area image (-888 mV and 500pA) showing  bright features surrounded  by standing-wave patterns, suggesting a strong modification of graphene layer as a result  of covalently grafted molecules. The bright features which show standing waves pattern are attributed to grafted molecules are highlighted in the insert in B). C) Zoomed image on small features for a sample immersed for 21 hours (-1.1V and 128 pA ). The insert in C) shows the Fourier Transform (FT) of the image. D) Schematic representation of the reciprocal lattice and possible scattering momentum vectors, representing zig-zag or arm-chair step edges, respectively. We represent the intravalley (momentum which connects two isocontours of same nature (K-K or K'-K') and intervalley scattering (for two different isocontour K-K') observed for zig-zag and armchair step edges, respectively.  Interference patterns observed for (E) a zig-zag step edge (-66.7mV, 121pA, 8.4x3.9nm2) and for (F) an armchair step edge (-27.3mV, 500pA, 8.4x3.9nm2).
}\label{Fig4}
\end{figure}

In figure \ref{Fig4}, we present  STM images of a typical sample. In A), a large image (400x400 nm$^{2}$) of a pristine and defect-free graphene layer,  as loaded in the UHV system, is shown together with a sample after  80 hours of immersion leading to the  deposition  of molecules on the graphene surface. We observe dark regions which correspond to the buffer layer. After immersion, we additionally observe many bright features, corresponding to the fluorinated maleimide molecules and potentially to some products introduced via the  rinsing process. Some of these deposited species were found to be highly mobile on the graphene surface, even at 77K, generating difficulties for obtaining proper STM images.  To verify that the images contained the signature of grafted molecules, we first scanned  at low bias (at a  few meV and large currents up to 1A). This allowed  to remove the potentially remaining free species.

\begin{figure}
\begin{center}
\includegraphics[width=5in]{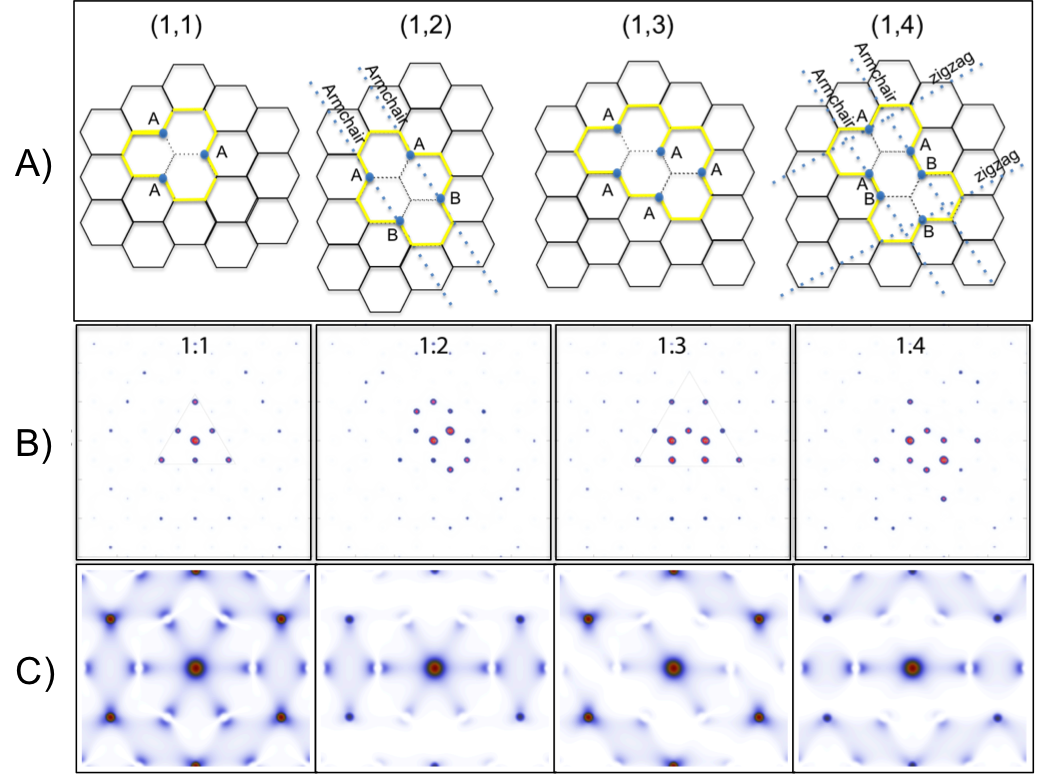}
\end{center}
\caption{A) Schematic representation of the possible defects created by the formation of $sp^{3}$ bonds. For a point defect configuration (1,1) and cycloaddition configurations (1,2), (1,3) and (1,4), respectively. T-matrix approximation calculations of the modified local density of states in direct space (B) and in K-momentum space (power spectrum) (C), respectively.}
\label{Fig5}
\end{figure}

After performing several scans and cleaning steps of the tip (achieved through high voltage pulses performed in areas outside the area of interest), we observed more stable bright features accompanied by standing-wave patterns, as shown in figure \ref{Fig4} B). These standing-wave patterns revealed a strong local modification of the underlying graphene lattice. The corresponding  $\sqrt{3}\times\sqrt{3}$ modulation of the density of states  is commonly observed around defects, generated, for example, by ion bombardment \cite{TapasztoPRB08},  or by  defects in the graphene layer (missing atoms), often decorated with impurities. These defects break the conjugation of the $sp^{2}$ bonds of graphene. This seems to be a necessary condition for the observation of strong standing-wave patterns. Indeed in a recent detailed STM study of nitrogen doping of graphene by substitution \cite{Lagoute2015}, standing-wave patterns were observed around N atoms substituting the C atoms only if vacancies were also present, such as for the pyridinic-N defect. In the latter case the conjugation is broken.  Quasi-particle interferences having a momentum-vector  which connects two different iso-energy contours around the K-point (K-K'), as described in the scheme shown in figure \ref{Fig4} D) are also called intervalley scattering \cite{SimonJPhsD11, CranneyPRB12}. In our case, for the smaller objects (see encircled object in the right lower part of figure \ref{Fig4} B)), the standing-wave patterns were  anisotropic with parallel  straight lines exhibiting a large contrast variation. Such  standing-wave patterns with large amplitudes are usually observed at armchair step edges, as shown in figure \ref{Fig4} F) \cite{Commentstepedges}. On the other hand patterns resulting from point defects or large defects (holes) typically  show a six-fold symmetry \cite{SimonEPJB2009, SimonJPhsD11}. An example of such pattern is observed around the  large object surrounded by arrows in figure \ref{Fig4} B) and exhibits a more isotropic standing wave pattern, which we attribute to a pile containing several molecules.
 Here, we discuss the possible standing-wave patterns created by a single molecule grafted onto the defect-free  surface of  graphene in analogy with the step-edge geometry. Figure \ref{Fig5} A) shows different configurations (n,p) for the cycloaddition reaction, where n and p represent the positions of the carbon atoms.  There,  grafting of a molecule changes a $sp^{2}$ bond into $sp^{3}$ bond. A (1,1) configuration corresponds to a single-site modification such as for a point defect, whereas a (1,p) configuration corresponds to a molecule grafted at two sites as expected for the D-A reaction of maleimide molecules. For the  (1,1) configuration, the electronic states of only one atom are perturbed (the molecule creates only one bond with the graphene surface). In the case of the formation of a $sp^{3}$ bond, the conjugation with the three first nearest neighbor bonds is broken (these affected bonds are represented by  dotted  lines). The resulting defect in the graphene lattice acts like a missing atom (a point defect for the (1,1) configuration) for which the three nearest neighbor atoms (represented as  blue dots in the figure), which belong to the sub-lattice A,  act as scatterers for the surrounding graphene quasiparticles. For the cycloaddition configuration (1,2), corresponding to grafting at  the first and  second carbon atom of a hexagonal cycle, the remaining scatterer atoms belong to both sub-lattices, i.e.,  A and B. If we consider the (1,2) $sp^{3}$ bonded atoms as missing atoms in the $sp^{2}$ network of graphene, the resulting defect is anisotropic and has a structure similar to that of an armchair-like edge. For a cycloaddition configuration (1,3), we achieve an isotropic defect with scatters atoms belonging only to the sub-lattice A. For the (1,4) cycloaddition configuration, the generated pattern becomes  more isotropic, however,  scatterers belong, as for the (1,2) configuration of the  cycloaddition reaction, to both sub-lattices, i.e.,  A and B. For the analysis of our experimental results, we thus focused on identifying the type of sub-lattice affected by the reaction,  because this allowed us to discriminate between patterns arising from  zigzag-like and armchair-like step edges.  Consequently, we are able to decide which type of defect symmetry (armchair-step-edge or zigzag-step-edge) was characteristic to the cycloaddition reaction and thus generated a characteristic pattern via the scattering and  interference process of quasiparticles.

\begin{figure}
\begin{center}
\includegraphics[width=6in]{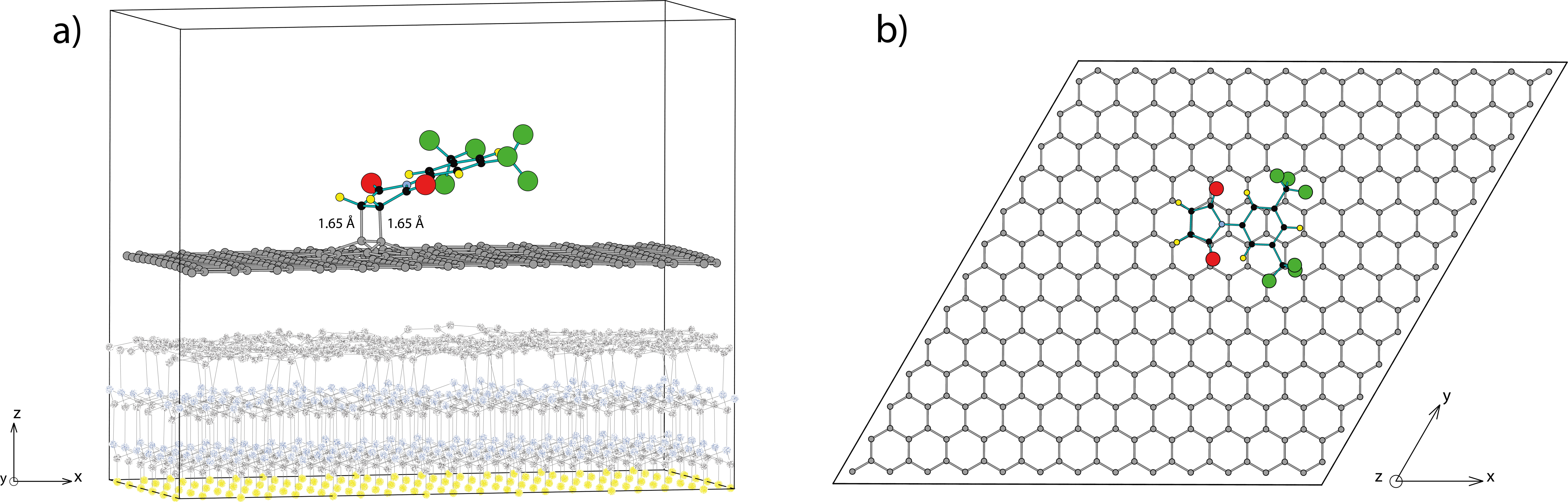}
\end{center}
\caption{ DFT optimized structure of the 1,2 cycloadduct for graphene on SiC. a) Side view of atomistic details of the supercell employed in the DFT calculations. b) Zoomed top view of a) }
\label{Fig6}
\end{figure}

In figures \ref{Fig6} B), we describe the modified electronic
structure due to impurity scattering for the impurity
configurations described in (figure \ref{Fig6} A)), respectively.
We use T-matrix approach to obtain the real space electron Green
function (GF) $G(r,r',\omega)$ . The local density of state $N(r)$
in real space is related to the GF through
$N(r)=-ImG(r,r',\omega)/\pi$. Only elastic scattering is
considered \cite{Jonas1, Jonas2,Jonas3,Jonas4}.
 We show in figure \ref{Fig6} C) the Fourier transforms (FT) of the local density of state calculated using the Born approximation, which is quasi-equivalent in this limit, to the k space T-matrix approximation \cite{SimonJPhsD11,Bena11}. For a (1,1)  point
defect, the modification of the the density of states exhibits a
clear six-fold symmetry. For the (1,2) configuration, the
calculation shows a rectangular-like anisotropic pattern generated
by the defect.   For the (1,3) configuration, we recover a
hexagonal  symmetry,  and for the (1,4) configuration, an
anisotropic pattern similar to, but less pronounced than the
pattern corresponding to the (1,2) configuration, the difference
consisting in that the (1,4) patterns recovers a hexagonal
symmetry at larger scale. These calculations show that, while the
shape of the defect is important,  the type of sub-lattice atoms
affected by the cycloaddition reaction is even more important.

 This can be  understood by analyzing  the patterns of the Fourier transforms (FT) of the direct space images (figure \ref{Fig6} C)) in comparison with the experimental FT shown in the insert figure \ref{Fig4}C). For the (1,1) configuration, representing a simple point defect, the FT  revealed strong features around the $\Gamma$ points, and less intense features around the K points with isotropic intensity (the intensity is the same for the six K-point features). A strong anisotropy is however observed for the (1,2) configuration of the cycloaddition reaction. The modified local density of states (LDOS) around the perturbed atoms shows a square-like anisotropic structure, and in the FT pattern,  this translates into the $\Gamma$-point intensity being less intense than for the (1,1) case and into a strong anisotropy between the six features around the K points. The reduction in the $\Gamma$-point intensity corresponds to the fact that in this case the intervalley scattering processes are dominant, similar to the standing-wave patterns observed for armchair step edges.

For the (1,3) configuration of the cycloaddition reaction, the FT pattern showed pronounced structures around the $\Gamma$ points and less intense features around the K points, as for the point defect (1,1). Here only scatterers form the A sub-lattice contributed which favors intervalley scattering, similar to the standing-wave patterns observed in the vicinity of zigzag step edges. For the last case, the (1,4) configuration of the cycloaddition reaction, same as for the (1,2) configuration of the cycloaddition reaction,  we observe a decrease of intensity for the features around the $\Gamma$ points, and an anisotropy in the features around the K points. We conclude that the (1,2) and (1,4) configurations of the cycloaddition reaction reveal  more anisotropic and more pronounced standing-wave patterns than the point defect (1,1) and the (1,3) configuration, and are thus the configurations consistent with the observed experimental features.\\
Finally, we have performed DFT calculations of interfacial properties (see THEORETICAL METHOD) in order to reveal the role of the substrate with respect to  forming a stable chemical bond between graphene and the molecules by a cycloaddition reaction. Our DFT calculations showed that both the (1,2) and (1,4) configuration  would allow a stable bond between   the molecule and a  free standing graphene layer. However, such a bond could not form for the (1,3) configuration. For graphene on SiC(0001), only the  (1,2) configuration resulted in a  stable bond, as depicted in Figure \ref{Fig6}.  The corresponding total energy of the resulting bond was  by ca. 0.4 eV lower as compared to the same bond  on free standing graphene. Using a SiC(0001) substrate is therefore favoring the  cycloaddition reaction. However, the (1,2) cycloaddition for graphene on SiC exhibits an energy level which is  by ca. + 2 eV higher than that of the reactant. This high energy value  would not normally allow  for the observed cycloaddition reaction, which, as we show,  was however possible at room temperature. Nevertheless we want to point out that we have performed this reaction not  in vacuum (as assumed in the calculations) but by immersing the graphene surface for about 80 hours in a toluene solution of molecules. This difference in the deposition conditions  hints at a lowering of the vacuum endothermicity due to the dissolution of the molecules by toluene.

\section{Conclusion}
The results of our study provide a strong evidence that cycloaddtion reaction between graphene and  fluorinated maleimide molecules is possible at room temperature. We observed that this reaction happened on a region of pristine graphene layer, which did not contain any pre-existing defect. The formation of covalent bonds was ascertained by the increase of the $sp^{3}$ component of the XPS spectrum. Furthermore, using STM we have visualized marked standing-wave patterns on  graphene, which were not observed for physisorbed molecules. These patterns, in particular  their geometry and symmetry, indicated that the  cycloaddition reaction occurred in the  (1,2) or (1,4) configuration, as confirmed by the T-matrix approximation. DFT calculations clearly show that (1,2) configuration is stable on G/SiC(0001). We have interpreted the  standing-wave patterns observed in direct space in analogy to standing-wave patterns observed in the vicinity of armchair and zig-zag step edges on graphene.  ARPES measurements revealed a tendency for the opening of  a gap and a slight decrease of the Fermi velocity. For bilayer graphene, the gap opening was more pronounced. Furthermore, we were able to distinguish which  sub-lattice of graphene was perturbed by  covalently grafting fluorinated maleimides molecules to graphene. In the future, in order to generate periodic molecular patterns chemically attached to  graphene, we will incorporate the reactive group used here in larger molecules which, in addition, allow for supramolecular self-assembly. We expect that, by creating covalent bonds, we will be able to form ribbons with a precise control of the nature (armchair or zigzag) of their edges.
\\
\\
\acknowledgments This work was supported by the R\'{e}gion Alsace,  the ANR ChimigraphN, the Universit\'e Franco-Allemande and the ERC Starting Independent Researcher Grant NANOGRAPHENE 256965.

\newpage

\begin{thebibliography}{35}
\bibitem{HaddonJACS2011}S. Sarkar, E. Bekyarova, S. Niyogi, R. C. Haddon, \emph{J. Am. Chem. Soc.}, 2011, \textbf{133}, 3324.
\bibitem{HaddonACR2012}S. Sarkar, E. Bekyarova, and R. C. Haddon, \emph{Accounts of Chemical Research}, 2012, \textbf{45}(4), 673.
\bibitem{Fukui1970}K. Fukui, Theory of Orientation and Stereoselection, \emph{Top. Curr. Chem.}, 1970, \textbf{15}(12), 1.
\bibitem{TownshendJACS1976}R. E. Townshend, G. Ramunni, G. Segal, G. W. J. Hehre, L. Salem, \emph{J. Am. Chem. Soc.}, 1976, \textbf{98}, 2190.
\bibitem{HoukJACS2013} Y. Cao, S. Osuna, Y. Liang, R. C. Haddon and K. N. Houk, \emph{J. Am. Chem. Soc}, 2013, \emph{135}(46),17643.
\bibitem{DenisChemEur2013} P.A. Pablo, \emph{Chemistry A European Journal}, 2013, \emph{19}, 15719.
\bibitem{Altenburg2015} S.J. Altenburg, M. Lattelais, B. Wang and M.-L. Bocquet and R. Berndt, \emph{J. Am. Chem. Soc.}, 2015, \textbf{137}, 9452.
\bibitem{Mattioli} Cristina MATTIOLI Thesis: "Design, synthesis and study of molecules for graphene functionalization". Chapter 6: "Maleimide derivatives for covalent functionalization of graphene". University of Toulouse.
\bibitem{Dujardin2015} M. Rubio-Roy, C. Mattioli, O. Couturaud, J.R. Huntzinger, A. Gourdon, E. Dujardin, \emph{submitted}.
\bibitem{simonPRB99}L. Simon, J. L. Bischoff, and L. Kubler, \emph{Phys. Rev. B}, 1999, \textbf{60}, 11653.
\bibitem{berger2002}C. Berger, Z. Song, T. Li, X. Li, A. Y. Ogbazghi, R. Feng, Z. Dai, A. N. Marchenkov, E. H. Conrad, P. N. First, and W. A. de Heer, \emph{J. Phys. Chem. B}, 2004, \textbf{108}, 19912.
\bibitem{kresse1996efficiency} G. Kresse and J. Furthm{\"u}ller, \emph{Computational Materials Science}, 1996, \textbf{6}, 15.
\bibitem{kresse1996efficient} G. Kresse and J. Furthm{\"u}ller, \emph{Phys. Rev. B},1996, \textbf{54}, 11169.
\bibitem{perdew1996generalized} J.P. Perdew, K. Burke and M. Ernzerhof, \emph{Phys. Rev. L}, 1996, \textbf{77}, 3865.
\bibitem{blochl1994projector} P.E. Bl{\"o}chl, \emph{Physical Review B}, 1994, \textbf{50}, 17953.
\bibitem{kresse1999ultrasoft} G. Kresse and D. Joubert, \emph{Phys. Rev. B}, 1999,  \textbf{59}, 1758.
\bibitem{riedl_structural_2007} C. Riedl, U. Starke, J. Bernhardt, M. Franke, and K. Heinz, \emph{Phys. Rev. B}, 2007, \textbf{76}, 245406.
\bibitem{charrier_solid-state_2002} A. Charrier, A. Coati, T. Argunova, F. Thibaudau, Y.Garreau, R. Pinchaux, I. Forbeaux, J.-M. Debever, M. Sauvage-Simkin and J.-M. Themlin, \emph{Journal of Applied Physics}, 2002, \textbf{92}, 2479.
\bibitem{brar_scanning_2007} V.W. Brar, Y. Zhang, Y. Yayon, A. Bostwick, T. Ohta, J.L. McChesney, K. Horn, E. Rotenberg, and M. F. Crommie, \emph{Applied Physics Letters}, 2007, \textbf{91}, 122102.
\bibitem{Emtsev2008}K. V. Emtsev, F. Speck, Th. Seyller, and L. Ley, \emph{Phys. Rev. B}, 2008, \textbf{77}, 155303.
\bibitem{RiedlJPhysD2010}C. Riedl C. Coletti, U. Starke, \emph{J. Phys. D: Appl. Phys.}, 2010, \textbf{43}, 374009.
\bibitem{TapasztoPRB08} L. Tapaszt\`o, G. Dobrik, P. Nemes-Incze, G. Vertesy, Ph. Lambin, and L. P. Bir\`o, \emph{Phys. Rev. B},2008, \textbf{78}, 233407.
\bibitem{Lagoute2015} Y. Tison, J. Lagoute, V. Repain, C; Chacon, Y. Girard, S. Rousset, F. Jouckert, D. Sharma, L. Henrard, H. Amara, A. Ghedjatti, and F. Ducastelle, \emph{ACSNano}, 2015, \textbf{9}, 670.
\bibitem{SimonJPhsD11}L. Simon, C. Bena, F. Vonau, M. Cranney, and D. Aubel, \emph{J. Phys. D},2011, \textbf{44}, 464010.
\bibitem{CranneyPRB12} M. Cranney, F. Vonau, P. B. Pillai, E. Denys, D. Aubel, M. M. De Souza, C. Bena, and L. Simon, \emph{Europhys. Lett.}, 2010, \textbf{91}, 66004.
\bibitem{SimonEPJB2009} L. Simon, C. Bena, F. Vonau, D. Aubel, H. Nasrallah, M. Habar and J.C. Peruchetti, \textmd{Eur. Phys. J. B}, 2009, \textbf{69}, 351.
\bibitem{Commentstepedges}Step edges play an important role in determining transport properties \cite{Sasaki10}, particularly in the case of ribbons of graphene, and are widely studied. Interpretation of the resulting edge electronic patterns are subject of intense debates, notably concerning the scattering processes from so-called zigzag or armchair edges \cite{DujardinNanolett, DujardinPNAS, MalletNanotech11}
\bibitem{Sasaki10} Ken-ichi Sasaki, Katsunori Wakabayashi, \emph{Phys. Rev. B},2010, \textbf{82}, 035421.
\bibitem{DujardinNanolett} H. Yang, A; J. Mayne, M. Boucherit, G. Comtet, G. Dujardin and Y. Kuk, \emph{Nano Lett.}, 2010, \textbf{10}, 943.
\bibitem{DujardinPNAS}C. Parka, H. Yangb, Andrew J. Mayne, G. Dujardin, S. Seod, Y. Kuka,
J. Ihma, and G. Kimd, \emph{PNAS}, 2011, \textbf{108}46, 18622.
\bibitem{MalletNanotech11} A. Mahmood, P. Mallet and J-Y Veuillen, \emph{Nanotechnology},2012, \textbf{23} 055706.
\bibitem{Jonas1} G. A. Fiete, E. J. Heller, \emph{Rev. Mod. Phys.}, 2003, \textbf{75}, 933.
\bibitem{Jonas2} J. Fransson and A. V. Balatsky, \emph{Phys. Rev. B},2012,  \emph{85} 161401R.
\bibitem{Jonas3} H. Hammar, P. Berggren and J. Fransson, \emph{Phys. Rev. B}, 2013, \textbf{88}, 254418.
\bibitem{Jonas4} J. Fransson, J-H. She, L. Pietronero, and A. V. Balatsky, \emph{Phys. Rev. B}, 2013,  \emph{87}, 245404.
\bibitem{Bena11}C. Bena and L. Simon Phys. Rev. B 83 (2011)115404.

\end{thebibliography}
\end{document}